\documentstyle[aps,prl,epsfig]{revtex}

\begin{document}
\draft

\twocolumn[\hsize\textwidth\columnwidth\hsize\csname @twocolumnfalse\endcsname

\title{Transverse momentum versus multiplicity fluctuations \\
in high-energy nuclear collisions }
\author{St. Mr\'owczy\'nski$^{1,2}$, M. Rybczy\'nski$^1$,
and Z. W\l odarczyk$^1$}

\address{$^1$Institute of Physics, \'Swi\c etokrzyska Academy,
ul. \'Swi\c etokrzyska 15, PL - 25-406 Kielce, Poland \\
$^2$So\l tan Institute for Nuclear Studies,
ul. Ho\.za 69, PL - 00-681 Warsaw, Poland}

\date{14-th September 2004}

\maketitle

\begin{abstract}

We discuss recently measured event-by-event fluctuations of transverse
momentum and of multiplicity in relativistic heavy-ion collisions.
It is shown that the non-monotonic behavior of the $p_T-$fluctuations
as a function of collision centrality can be fully explained by
the observed non-monotonic multiplicity fluctuations. A possible
mechanism responsible for the multiplicity fluctuations is also 
considered.

\end{abstract}

\vspace{0.2cm}

%\pacs{25.75.-q, 25.75.Gz}

%{\it Keywords:} Relativistic heavy-ion collisions, fluctuations

\vspace{0.2cm}

]

Event-by-event fluctuations of transverse momentum in heavy-ion collisions
have been recently measured both at CERN SPS
\cite{Appelshauser:1999ft,Anticic:2003fd,Adamova:2003pz}
and BNL RHIC \cite{Adcox:2002pa,Adler:2003xq,Voloshin:2001ei,Adams:2003uw,Pruneau:2004nc},
see also a brief review \cite{Mitchell:2004xz}. The data, which show
a non-trivial behavior as a function of collision centrality, have 
been theoretically discussed from very different points of view
\cite{Gazdzicki:1997fg,Liu:1998xf,Bleicher:1998wu,Mrowczynski:1998vt,Stephanov:1999zu,Capella:1999uc,Baym:1999up,Dumitru:2000in,Korus:2001fv,Korus:2001au,Stephanov:2001zj,Voloshin:2002ku,Gavin:2003cb,DiasdeDeus:2003sn,Rybczynski:2003jk,Ferreiro:2003dw,Liu:2003jf,Stephanov:2004wx},
including complete or partial equilibration \cite{Bleicher:1998wu,Mrowczynski:1998vt,Stephanov:2001zj,Gavin:2003cb}, critical phenomena 
\cite{Stephanov:1999zu,Stephanov:2004wx}, string or cluster percolation 
\cite{DiasdeDeus:2003sn,Ferreiro:2003dw}, production of jets 
\cite{Liu:1998xf,Liu:2003jf}. In spite of these efforts, a mechanism 
responsible for the fluctuations is far from being uniquely identified.
Recently, the NA49 Collaboration published very first data on multiplicity 
fluctuations as a function of collision centrality 
\cite{Gazdzicki:2004ef,Rybczy2004}.
Unexpectedly, the ratio $Var(N)/\langle N \rangle$, where $Var(N)$ is the 
variance and $\langle N \rangle$ is the average multiplicity of negative 
particles, changes non-monotonically when number of wounded nucleons\footnote{
A nucleon is called wounded if it interacts at least once in the course of 
nucleus-nucleus collision. The number of wounded nucleons $N_w$ approximately 
equals the number of participants, and we assume here that the equality 
holds.} grows. It is close to unity at fully peripheral ($N_w \le 10$) 
and completely central ($N_w \ge 250$) collisions but it manifests 
a prominent peak at $N_w \approx 70$, as shown in Fig.~1a. The measurement 
has been performed at the collision energy 158-A-GeV in the transverse 
momentum and pion rapidity intervals $(0.005,1.5)$ GeV and $(4.0,5.5)$,
respectively. The azimuthal acceptance has been also limited, and about 
$20\%$ of all produced negative particles have been used in the analysis.

The aim of this note is to show that the non-trivial behavior of transverse
momentum fluctuations can be explained by the multiplicity fluctuations 
which enter the measures of $p_T-$fluctuations. Specifically, we assume 
that in nucleus-nucleus collisions the event's transverse momentum 
is correlated to event's multiplicity exactly as in the proton-proton 
interactions \cite{Kafka:1976py}, and we express the $\Phi-$measure
\cite{Gazdzicki:ri} of $p_T-$fluctuations through the multiplicity 
fluctuations. It is convenient for our discussion to use data on the 
transverse momentum and multiplicity fluctuations measured in the same 
experimental conditions. For this reason, we choose the data obtained 
by the NA49 collaboration which used the $\Phi-$measure \cite{Gazdzicki:ri} 
to quantify the fluctuations of transverse momentum. 

Let us first introduce the measure. One defines the single-particle 
variable $z \buildrel \rm def \over = x - \overline{x}$ with the overline
denoting averaging over a single particle inclusive distribution. 
Here, we identify $x$ with the particle transverse momentum $p_T$. The 
event variable $Z$, which is a multiparticle analog of $z$, is defined as
$Z \buildrel \rm def \over = \sum_{i=1}^{N}(x_i - \overline{x})$, where
the summation runs over particles from a given event. By construction,
$\langle Z \rangle = 0$ where $\langle ... \rangle$ represents averaging
over events. Finally, the $\Phi-$measure is defined in the following way
$$
\Phi (x) \buildrel \rm def \over =
\sqrt{\langle Z^2 \rangle \over \langle N \rangle} -
\sqrt{\overline{z^2}} \;.
$$
It is evident that $\Phi = 0$, when no inter-particle correlations are 
present. Consequently, $\Phi$ is `deaf' to the statistical noise. The 
measure also possesses a less trivial property. Namely, $\Phi$ is 
{\it independent} of the distribution of number of particle sources 
if the sources are identical and independent from each other 
\cite{Gazdzicki:ri,Mrowczynski:1999un}. Thus, the $\Phi-$measure is 
`blind' to the impact parameter variation as long as the `physics' does 
not change with the collision centrality. In particular, the $\Phi$ is
independent of the impact parameter if the nucleus-nucleus collision is 
a simple superposition of nucleon-nucleon interactions.

$\Phi (p_T)$ measured in nucleus-nucleus collisions at SPS energy as 
a function of centrality \cite{Anticic:2003fd} is shown in Fig.~1b. 
The measurement has been performed in exactly the same experimental 
conditions as that of multiplicity fluctuations shown in Fig.~1a. 
As seen, both transverse momentum fluctuations expressed in terms
of $\Phi$ and multiplicity fluctuations display a very similar 
centrality dependence, suggesting that they are related to each 
other. 

In the very first paper, where the $\Phi-$measure was introduced
\cite{Gazdzicki:ri}, it was argued that the correlation between the 
event's multiplicity and transverse momentum is a main source of the 
$p_T-$fluctuations as quantified by $\Phi$. For the case of p-p interactions, 
the problem was then studied in detail in \cite{Korus:2001fv}. Following 
this paper, we introduce the correlation $\langle p_T \rangle$ {\it vs.} $N$ 
through the multiplicity dependent temperature or slope parameter of 
$p_T-$distribution. Specifically, the single particle transverse momentum 
distribution in the events of multiplicity $N$ is chosen in the form
suggested by the thermal model {\it i.e.}
\begin{equation}\label{pt-dis}
P_{(N)}(p_T) \sim p_T
\exp \bigg[ - {\sqrt{m^2 + p_T^2} \over T_N} \bigg] \;,
\end{equation}
where $m$ is the particle mass while $T_N$ is the multiplicity
dependent temperature. In Ref. \cite{Korus:2001fv}, $T_N$ was defined as
\begin{equation}\label{T-N-old}
T_N = T + \Delta \! T \, ( \langle N \rangle - N )
\end{equation}
with $\Delta \! T$ controlling the correlation strength. The 
parameterization (\ref{T-N-old}) was reasonable for proton-proton 
collisions where $\langle N \rangle$ is fixed, but it is not 
reasonable to study the centrality dependence in A-A collisions
where $\langle N \rangle$ varies.

\vspace{-3.3cm}
\begin{figure}
\centerline{\epsfig{file=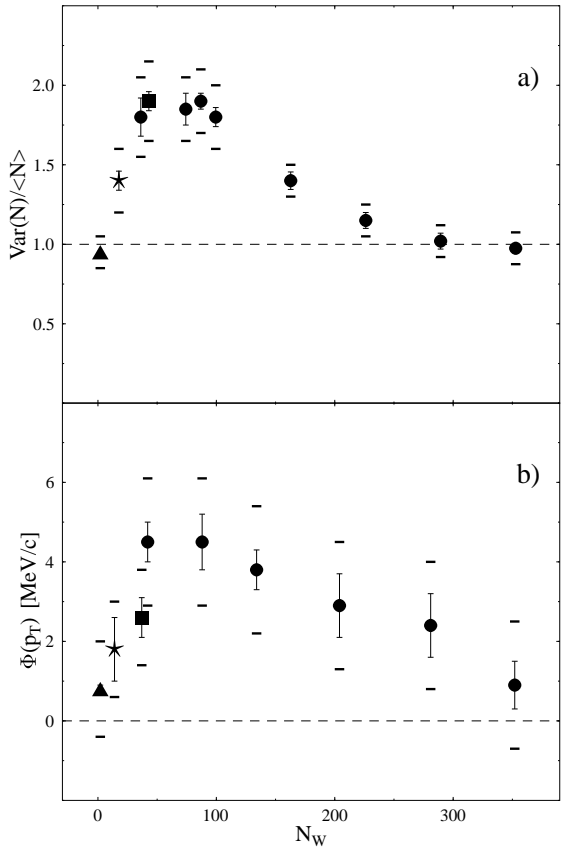,width=1.1\linewidth}}
\vspace{-0.3cm}
\caption{Multiplicity (a) and transverse momentum (b) fluctuations
of negative particles as a function of number of wounded nucleons.
The triangles correspond to p-p collisions, asterisks to C-C, squares
to Si-Si, and circles to Pb-Pb. There are denoted statistical errors
with vertical bars and total errors including systematic uncertainties
with horizontal dashes. The multiplicity data are taken from
\protect\cite{Gazdzicki:2004ef,Rybczy2004} while those on the transverse
momentum from \protect\cite{Anticic:2003fd}.}
\end{figure}

The correlation $\langle p_T \rangle$ {\it vs.} $N$ at SPS energy, which 
is directly observed \cite{Kafka:1976py,Anticic:2003fd} in p-p collisions, 
is most probably of simple kinematical origin. Namely, when the 
multiplicity of produced particles grows at fixed collision 
energy, there is less and less energy to be distributed among transverse 
degrees of freedom of produced particles. Consequently, the average 
event's $p_T$ decreases when $N$ grows. We expect that the correlation 
$\langle p_T \rangle$ {\it vs.} $N$ is also present in A-A collisions
at fixed centrality as the number of wounded nucleons controls the 
amount of energy to be used for particle production. However, we replace 
the parameterization (\ref{T-N-old}) by 
\begin{equation}\label{T-N}
T_N = T + \delta T \bigg(1 - \frac{N}{\langle N \rangle}\bigg) \;,
\end{equation}
with $\delta T = \Delta \! T \langle N \rangle$. The relation (\ref{T-N}) 
correlates the slope parameter $T_N$ to the event's multiplicity $N$
at fixed $\langle N \rangle$. The parameters $T$ and $\delta T$ are 
assumed to be independent of the centrality while the average multiplicity 
$\langle N \rangle$ depends (roughly linearly) on $N_w$. As will be seen 
in our final formula (\ref{final}), a small variation of $T$ with the 
centrality does not much matter.

The inclusive transverse momentum distribution, which determines
$\overline{z^2}= \overline{p^2_T} -\overline{p_T}^2$, reads
$$
P_{\rm incl}(p_T) = {1 \over \langle N \rangle}
\sum_N {\cal P}_N N P_{(N)}(p_T) \;,
$$
where ${\cal P}_N$ is the multiplicity distribution. The $N-$particle 
transverse momentum distribution in the events of multiplicity $N$ is 
assumed to be the $N-$product of $P_{(N)}(p_T)$. Therefore, all 
inter-particle correlations different than $\langle p_T \rangle$ {\it vs.} 
$N$ are neglected here.  Then, one easily finds
\begin{eqnarray*}
\langle Z^2 \rangle &=& \sum_N {\cal P}_N
\int_0^{\infty} dp_T^1 .\; . \; . \int_0^{\infty} dp_T^N
\\
&\times&
\Big(p_T^1 +  .\; . \; . + p_T^N - N \overline{p_T} \Big)^2
P_{(N)}(p_T^1) \; . \; . \; . \; P_{(N)}(p_T^N) \;.
\end{eqnarray*}
Assuming that the particles are massless and the correlation is weak,
{\it i.e.} $T \gg \delta T$, the calculation of $\Phi$ can be 
performed analytically. The result is \cite{Korus:2001fv}
\begin{eqnarray}
\label{massless-phi}
\Phi(p_T) &=& \sqrt{2}\, {(\delta T)^2 \over T \langle N \rangle^5}
\big( \langle N^4 \rangle \langle N \rangle^2
   -2 \langle N^3 \rangle \langle N^2\rangle \langle N \rangle
\\[1mm] \nonumber
&& - \: \langle N^3 \rangle \langle N \rangle^2
    + \langle N^2 \rangle^3
    + \langle N^2 \rangle^2 \langle N \rangle \big) \;,
\end{eqnarray}
where terms of the third and higher powers of $\delta T$ have
been neglected. As seen, the lowest non-vanishing contribution to 
$\Phi$ is of the second order in $\delta T$.

We intend to express $\Phi (p_T)$ through $Var(N)/\langle N \rangle$
but $\Phi (p_T)$, as given by Eq.~(\ref{massless-phi}), also depends
on the third and fourth moment of multiplicity distribution. It would 
be in the spirit of our minimalist approach to use the multiplicity 
distribution which maximizes the Shannon's information entropy 
$S \equiv \sum_N {\cal P}_N \, {\rm ln} {\cal P}_N$ \cite{Shanon48} with 
$\langle N \rangle$ and $Var(N)$ being fixed. The least biased method to 
obtain statistical distribution was prompted by Jaynes \cite{Jaynes57}.
An application of the information theory to phenomenology of high-energy 
collisions is discussed in \cite{Wilk:1990bf}. The multiplicity distribution, 
which maximizes the entropy at fixed $\langle N \rangle$ and $Var(N)$, is 
given by the formula
\begin{equation}\label{minimal}
{\cal P}_N = {\rm exp}\Big(a + b\, N + c\, N^2 \Big) \;,
\end{equation}
where the parameters $a$, $b$ and $c$ are determined by the equations
$$
\sum_N {\cal P}_N = 1 \;,\;\;\;\;\;
\sum_N N {\cal P}_N = \langle N \rangle \;,
$$
\vspace{-5mm}
$$
\sum_N (N-\langle N \rangle )^2 {\cal P}_N = Var(N) \;.
$$ 
Unfortunately, there are no simple analytic expressions of $a$, $b$ 
and $c$, and consequently the distribution (\ref{minimal}) is very
inconvenient to use. However, under the condition
$\langle N \rangle \gg \sqrt{Var(N)} \gg 1$, which is usually satisfied 
in A-A collisions at fixed centrality, the distribution (\ref{minimal}) 
can be replaced by the continuous Gauss distribution. Then, we get the 
required relations: $\langle (N-\langle N \rangle )^3 \rangle = 0$ and 
$ \langle (N-\langle N \rangle )^4 \rangle
= 3 \langle (N-\langle N \rangle )^2 \rangle^2$.

\vspace{-1.2cm}
\begin{figure}
\centerline{\epsfig{file=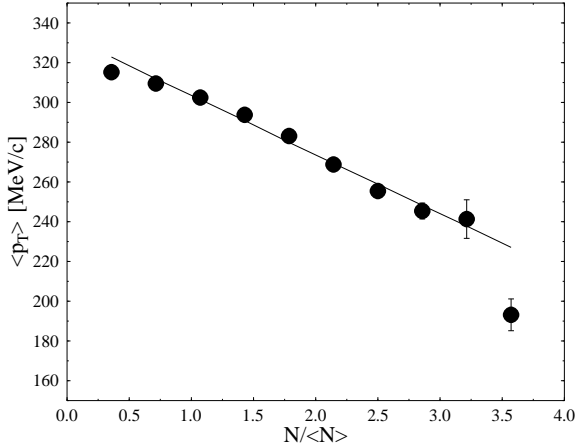,width=1.1\linewidth}}
\vspace{-0.3cm}
\caption{The average transverse momentum of negatively charged particles
produced in p-p collisions as a function of event's negative particle
multiplicity divided by the mean. The data are taken from
\protect\cite{Anticic:2003fd} where the acceptance is precisely defined.
The line corresponds to $T=137$ MeV and $\delta T= 15.5$ MeV.}
\end{figure}

Using these relations, the expression (\ref{massless-phi}) gets 
the form
\begin{eqnarray*}
\Phi(p_T) &=& \sqrt{2} \: \frac{(\delta T)^2}{T} \;
\frac{Var(N)}{\langle N \rangle}
\\[1mm]
&\times&
\bigg[1 - \frac{1}{\langle N \rangle}
+ \frac{Var^2(N)}{\langle N \rangle^4}
+ \frac{Var(N)}{\langle N \rangle^3} \bigg] \;.
\end{eqnarray*}
Taking into account the already adopted assumption that
$\langle N \rangle \gg \sqrt{Var(N)} \gg 1$, we finally find
\begin{equation} \label{final}
\Phi(p_T) = \sqrt{2} \: \frac{(\delta T)^2}{T} \;
\frac{Var(N)}{\langle N \rangle} \;.
\end{equation}
When the Negative Binomial Distribution, instead of the Gauss, is used 
to describe the multiplicity distribution, one obtains the formula, 
which in the limit $\langle N \rangle \gg \sqrt{Var(N)} \gg 1$, coincides 
with Eq.~(\ref{final}).

The values of the parameters $T$ and $\delta T$ for p-p collisions 
can be obtained from the NA49 data published in \cite{Anticic:2003fd}. 
Following \cite{Korus:2001fv}, we have computed the average $p_T$ at fixed 
$N$, using the distribution (\ref{pt-dis}) with $T_N$ given by Eq.~(\ref{T-N}). 
Comparing the results of our calculations with the experimental data 
\cite{Anticic:2003fd}, which are shown in Fig.~2, we have found $T = 137$ 
MeV and $\delta T= 15.5$ MeV. We note that $T$ and $\delta T$ are 
essentially independent from each other when the experimental data are fitted
as $\delta T$ determines the slope of the curve shown in Fig.~2 while $T$ 
controls its vertical position. For $T = 137$ MeV and $\delta T= 15.5$ MeV,
the coefficient in the formula (\ref{final}) equals
\begin{equation}
\label{coef}
\sqrt{2} \: \frac{(\delta T)^2}{T} \approx 2.48 \;\;{\rm MeV} \;.
\end{equation}

In Fig.~3 we compare the experimental values of $\Phi (p_T)$ with the
predictions of the formula (\ref{final}) with the numerical coefficient
given by Eq.~(\ref{coef}). As seen, the agreement is quite 
satisfactory\footnote{In Ref. \cite{Korus:2001fv} the parameters $T$ and 
$\delta T$ were estimated as 167 and 8.2 MeV, using the data on p-p collisions 
at 205 GeV \cite{Kafka:1976py}. The data \cite{Anticic:2003fd} were not 
available at that time. Then, $\sqrt{2} (\delta T)^2/T \approx 0.57$,
and the value of $\Phi (p_T)$ calculated by means of the formula (\ref{final})
is dramatically underestimated. It was also concluded in \cite{Korus:2001fv} 
that the $p_T$ {\it vs.} $N$ correlation produces too small value of 
$\Phi (p_T)$ to explain the experimental value. Now, this conclusion must 
be revoked. The discrepancy between the data \cite{Kafka:1976py} and 
\cite{Anticic:2003fd} can be easily explained. The data \cite{Kafka:1976py} 
were collected at higher collision energy in the full phase-space while the 
NA49 measurement \cite{Anticic:2003fd} was performed, as already mentioned, 
in the forward rapidity window (4, 5.5).}. However, the analytic result 
(\ref{final}) has been derived under several rather rough approximations. 
Therefore, we have also performed a Monte Carlo simulation which is free 
of these approximations. For every nucleus-nucleus collision at a given 
centrality, we have first generated its multiplicity, using the Negative 
Binomial Distribution with the mean value and variance as in the experimental 
data \cite{Gazdzicki:2004ef,Rybczy2004}. Further, we have attributed the 
transverse momentum from the distribution (\ref{pt-dis}) with $T = 137$ MeV 
and $\delta T= 15.5$ MeV to each particle assuming, as in the experimental 
analysis \cite{Anticic:2003fd}, that all particles are pions. Having 
a sample of events for every centrality, the $\Phi-$measure has been 
computed. The statistical errors have been determined by means of the 
subsample method. The results of our simulation are confronted with the 
experimental data in Fig.~3. As seen, there is a perfect agreement.

Our calculations explicitly take into account only the $p_T$ {\it vs.}
$N$ correlation. However, other correlations, in particular those due
to quantum statistics, are not entirely neglected. Since we use
the experimental value of $Var(N)/\langle N \rangle$, all correlations
which contribute to this quantity, are effectively taken into account
in our estimate of $\Phi (p_T)$.

The multiplicity fluctuations strongly influence the $p_T-$fluctuations
expressed in terms of $\Phi$, as the measure $\Phi$ depends on the
particle multiplicity distribution. It should be stressed that other
fluctuations measures, such as $F$ used by PHENIX Collaboration 
\cite{Adcox:2002pa,Adler:2003xq} or $\Delta \sigma$ and $\sigma_{\rm dyn}$
by STAR \cite{Voloshin:2001ei,Adams:2003uw}, are also influenced 
by multiplicity fluctuations. Therefore, our main result (\ref{final}) 
can be easily translated for $F$, $\Delta \sigma$ or $\sigma_{\rm dyn}$.

\vspace{-1.2cm}
\begin{figure}
\centerline{\epsfig{file=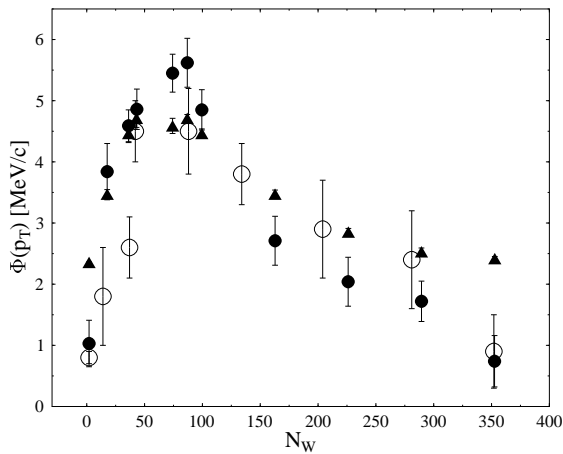,width=1.1\linewidth}}
\vspace{-0.3cm}
\caption{$\Phi(p_T)$ as a function of number of wounded nucleons.
The open circles correspond to the NA49 data \protect\cite{Anticic:2003fd},
the full circles show the results of our simulation while the triangles
present the prediction of the analytical formula (\protect\ref{final})
with the numerical coefficient given by Eq.~(\protect\ref{coef}).}
\end{figure}

A specific pattern of the $p_T-$fluctuations have been explained by
the observed multiplicity fluctuations. Before closing our considerations
we briefly consider a possible origin of the non-monotonic dependence
of $Var(N)/\langle N \rangle$ on the collision centrality. For this
purpose we first derive a well-known formula which relates particle 
number fluctuations to inter-particle correlations. The average 
multiplicity can be written as
$$
\langle N \rangle = \int_V d^3r \: \rho ({\bf r}) \;.
$$
$V$ is the volume of the interaction zone (fireball), where the 
particles are produced, and $\rho ({\bf r})$ is the particle density. 
The average multiplicity of produced particles is known to be roughly 
proportional to the number of wounded nucleons $N_w$ \cite{Afanasiev:2002mx}. 
Since $N_w$ is in turn proportional to the volume $V$, we have 
$\langle N \rangle = \bar\rho \: V$ with $\bar\rho$ being constant. 
The second moment of the multiplicity distribution can be written as 
$$ 
\langle N (N - 1) \rangle =  \int_V d^3r_1 \int_V d^3r_2 \: 
\rho_2 ({\bf r}_1, {\bf r}_2)  \;,
$$
where $\rho_2 ({\bf r}_1, {\bf r}_2)$ is the two-particle density.
Defining the correlation function $\nu ({\bf r}_1 - {\bf r}_2)$ 
through the equation
$$
\rho_2 ({\bf r}_1, {\bf r}_2) = \rho ({\bf r}_1) \: \rho ({\bf r}_2)
\Big( 1 + \nu ({\bf r}_1 - {\bf r}_2) \Big) \;,
$$
we get the desired formula
\begin{equation} 
\label{var-corr1}
\frac{Var(N)}{\langle N \rangle} = 1 +
\bar\rho  \int_V d^3r \:\nu ({\bf r}) \;,
\end{equation}
which tells us that the multiplicity distribution is poissonian
if particles are independent from each other ($\nu ({\bf r})$ = 0). 
The pattern seen in Fig.~1a clearly shows that the particles are 
correlated at the stage of production. 

For further discussion we assume that the fireball is spherically 
symmetric and that its radius equals $R \approx r_0 N_w^{1/3}$ with
$r_0 \approx 1 \; {\rm fm}$. Then, the formula (\ref{var-corr1}) reads
\begin{equation}
\label{var-corr2}
\frac{Var(N)}{\langle N \rangle} = 1 + 4\pi \:
\bar\rho  \int_0^R dr\, r^2 \, \nu (r) \;.
\end{equation}
It is not difficult to invent a correlation function $\nu (r)$ which 
substituted in Eq.~(\ref{var-corr2}) reproduces the data shown in Fig.~1a. 
Various functions are discussed in \cite{Rybczynski:2004zi}. Here we only 
describe qualitative features of $\nu (r)$. The correlation function has 
to be positive at small distances (attractive interaction) and negative 
at larger ones (repulsive interaction). The sign of the correlation changes
at $r \approx 4 \; {\rm fm}$ which corresponds to $N_w \approx 70$
when $Var(N)/\langle N \rangle$ reaches its maximum. For 
$r \gtrsim (300)^{1/3} \approx 7 \; {\rm fm}$ the correlation
function vanishes. A physical mechanism responsible for such
a correlation function is rather unclear but some possibilities,
which include: combination of strong and electromagnetic interaction,
percolation, dipole-dipole interaction, and non-extensive thermodynamics,
are discussed in \cite{Rybczynski:2004zi}.

We conclude our considerations as follows. A non-trivial behavior of 
transverse momentum fluctuations as a function of collision centrality 
can be fully explained by the centrality dependence of multiplicity 
fluctuations if the mean transverse momentum is correlated to the particle 
multiplicity in nucleus-nucleus collisions as in the proton-proton 
interactions. This correlation is most probably of simple kinematic origin. 
Our observation seem to exclude various exotic explanations of transverse 
momentum fluctuations. However, a mechanism responsible for multiplicity 
fluctuations still needs to be clarified. 

%**************************************************************|
\acknowledgements
%**************************************************************|

We are very grateful to Kasia Grebieszkow, Marek Ga\'zdzicki
and Peter Seyboth for fruitful discussions.

\end{document}